# Discovery of novel magnetic Y-Mn-B compounds via advanced machine learning guided framework

Weiyi Xia[1,2], Wei Shen Tee[2], Maxim Moraru[3], Ying Wai Li[3], Cai-Zhuang Wang[1,2,*]

[1]*Ames National Laboratory, U.S. Department of Energy, Iowa State University, Ames, Iowa 50011, USA*

[2]*Department of Physics and Astronomy, Iowa State University, Ames, Iowa 50011, USA*

[3]*Computer, Computational, and Statistical Sciences Division, Los Alamos National Laboratory, Los Alamos, NM 87545, USA*

## Abstract

Rare-earth transition-metal borides offer critical structural motifs for permanent-magnet design; however, the manganese-rich regions within these compositional phase spaces remain largely unexplored. In this work, we develop an advanced machine-learning-assisted discovery framework to explore Y-Mn-B ternary system. Starting from over one million hypothetical structures generated from known structures in databases, we filtered promising candidates by first applying graph neural networks to predict material stability, then using machine-learning-interatomic-potential to relax their structures, and finally validating the results with first-principles calculations. We identify 5 stable and near-stable Y-Mn-B phases along with 61 metastable compounds with the formation energy within 100 meV/atom with respect to the ternary convex hull. Among them, $Y_2Mn_7B_7$ and $YMn_4B_4$ are structurally analogous to the previously synthesized $R_{1+\varepsilon}Fe_4B_4$ 1D incommensurate composite chain compounds. In striking contrast to the strongly suppressed Fe moments reported, our first-principles calculations reveal that the predicted Mn-chain phases preserve sizable local Mn moments ($\approx 1.1\ \mu_B$) and favored ferromagnetic ordering. Electronic structure analyses elucidate the microscopic origin of moment recovery via an enhanced exchange splitting driven by a Stoner-like instability. We also perform systematic Mn-Fe substitution to confirm the thermodynamic continuity and a monotonic enhancement of the macroscopic magnetization, from Fe to Mn. These findings indicate that targeted transition-metal substitution within a one-dimensional boride family can recover transition-metal magnetism, offering a physically interpretable route for designing new magnetic rare-earth transition-metal borides.

## 1. Introduction

Permanent magnets are vital to modern energy conversion and data storage, driving innovations from electric vehicles to magnetic recording media. Driven by the surging demand for high-performance components, rare-earth transition-metal compounds remain at the forefront of magnetic materials design. In the archetypal $R_2Fe_{14}B$ family, the transition-metal sublattice supplies most of the saturation magnetization, whereas the rare-earth sublattice provides the spin-orbit-driven magnetocrystalline anisotropy required to resist demagnetization. This separation of roles is especially clear in $Nd_2Fe_{14}B$, where site-specific experiments and first-principles calculations show that rare-earth crystal-field physics controls much of the magnetocrystalline anisotropy, even though the Fe network dominates the total moment [1-3]. Yttrium-containing analogs such as $Y_2Fe_{14}B$ are therefore useful non-4*f* reference systems: they retain the transition-metal-boride framework while removing localized rare-earth 4*f* contributions, allowing the intrinsic magnetic behavior of the transition-metal network to be examined more directly [2,4].

In the R-Fe-B system, besides the well-known 2-14-1 phase, a boron-rich family $R_{1+\varepsilon}Fe_4B_4$ has also attracted a lot of attention. These compounds occur as secondary or grain-boundary-related phases in Nd-Fe-B materials and have been the subject of sustained crystallographic and magnetic study [5-7]. Structurally, they consist of two interpenetrating substructures: strings of rare-earth atoms extending along the c axis and Fe-B networks containing chains of edge-sharing Fe tetrahedra with associated boron dimers [5,8]. Because the two substructures have different repeat lengths along the chain direction, these phases are often described either as incommensurate composite structures or as commensurate approximants, including the 1-4-4, 5-18-18, and 2-7-7 compositions. The 2-7-7 composition has been reported for $Nd_2Fe_7B_7$ in the *Pccn* space group [7,9-10].

Magnetically, the Fe-based 1-4-4 family suffers from the severe suppression of the iron moment. Early single-crystal and Mössbauer studies showed that $Nd_5Fe_{18}B_{18}$ (or $Nd_{1.11}Fe_4B_4$) hosts essentially nonmagnetic Fe atoms. Consequently, its low-temperature ferromagnetic order stems almost exclusively from the rare-earth sublattice [9]. A systematic Mossbauer spectroscopy and magnetization study of $R_{1+\varepsilon}Fe_4B_4$ similarly found low ordering temperatures across the rare-earth series and concluded that the Fe ion carries zero magnetic moment within experimental sensitivity [11]. More recent single-crystal work on $Nd_{1+\varepsilon}Fe_4B_4$ confirmed the low-temperature character of the magnetic order and emphasized the complexity of the modulated structure and anisotropy [7]. Thus, although the $R_{1+\varepsilon}Fe_4B_4$ architecture is structurally proximate to high-performance magnetic phases, it does not provide the strong transition-metal magnetization needed for permanent-magnet applications.

A natural route to altering this stability-magnetization balance is to replace Fe by another magnetic 3*d* transition metal, such as Mn. However, prior studies of Mn substitution in the 2-14-1 framework show that such a strategy is not straightforward. Mossbauer spectroscopy and neutron diffraction measurements on $Y_2(Fe_{1-x}Mn_x)_{14}B$ revealed that Mn preferentially occupies specific transition-metal sites, especially large-volume Fe sites. The study also identified that Mn favors antiferromagnetic coupling in this environment [12]. Magnetic measurements on $R_2(Fe_{1-x}Mn_x)_{14}B$ with R = Nd, Pr, and Y showed that increasing the Mn concentration drastically reduces both the saturation magnetization and the Curie temperature, a behavior attributed to enhanced antiferromagnetic coupling [13]. These results indicate that simple substitution is not an effective strategy for enhancing ferromagnetism: although Mn can introduce large local moments, it can also generate competing exchange interactions that suppress macroscopic ferromagnetism.

This motivates a direct search of the Y-Mn-B ternary system rather than treating Mn only as a dopant in known Fe-boride frameworks. The relevant question is whether this Mn-B motif can stabilize these moments and align them coherently enough to form a useful magnetic sublattice. Yttrium is an excellent reference element for this purpose because it removes localized rare-earth 4*f* physics while retaining the size and spacer role of a rare-earth-like atom.

In this work, we apply an advanced version of exa-AMD, a machine-learning-assisted framework for the accelerated discovery and design of novel materials [20], to explore the ternary Y-Mn-B system for novel magnetic materials. Related ML-guided workflows have already been used to discover magnetic and ternary compounds across chemically complex spaces [14-19]. Our present study targets a more specific structure-property question: can a Y-Mn-B compound be thermodynamic stabilized while maintaining a substantial magnetization, without the antiferromagnetism often encountered when Mn is introduced into known Fe-boride magnets? To address this question, we first introduce the exa-AMD framework, together with an advanced machine-learning interatomic potential (MLIP)-based relaxation and convex-hull sorting step, and then apply it for a comprehensive search of Y-Mn-B ternary compounds. Our study identifies 5 stable and near-stable Y-Mn-B phases, along with 61 metastable compounds. Among them, the stable $Y_2Mn_7B_7$ phase and the near-stable $YMn_4B_4$ approximant share a one-dimensional chain motif related to the previously synthesized $R_{1+\varepsilon}Fe_4B_4$ family, but they retain substantial Mn-derived magnetization. Therefore, we use this non-4*f* system to decouple transition-metal magnetism from rare-earth anisotropy. We demonstrate that the same boride-chain architecture that suppresses Fe moments can host an active Mn magnetic sublattice when the local Mn-B bonding and nodal electron concentration are appropriately tuned.

# 2. Methods

## 2.1 exa-AMD: A ML-guided high-throughput discovery workflow

The workflow consisted of five main stages: structure generation, ML formation-energy prediction, structure filtering, MLIP relaxation and hull-prioritized sorting, and DFT calculation. A schematic of the ML-guided exa-AMD workflow for searching ternary Y-Mn-B compounds is shown in Fig. 1.

The workflow is implemented in exa-AMD, a Python application that integrates machine-learning models with first-principles calculations [20]. Parsl manages the parallel execution of CPU and GPU tasks [21]. The modular implementation permits calculations to restart from intermediate stages and allows computing resources to be assigned according to the requirements of each task.

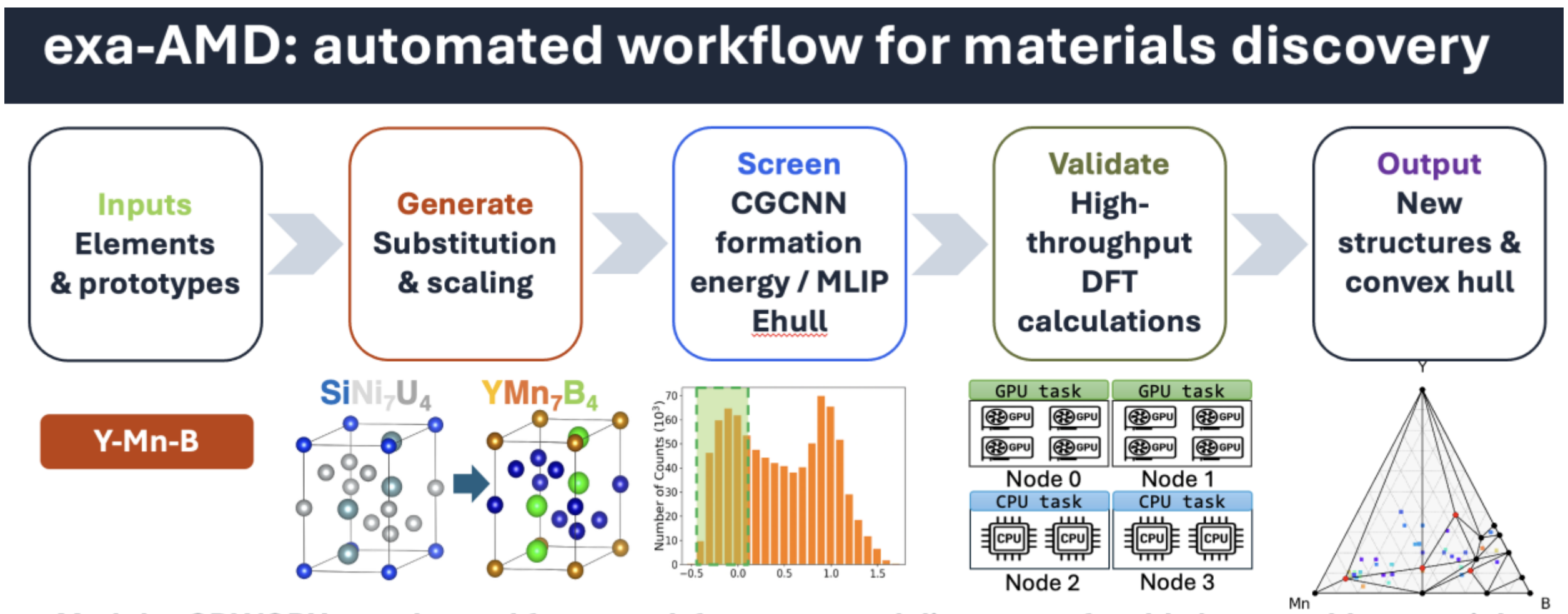

**Figure 1.** Automated exa-AMD workflow for Y-Mn-B materials discovery. The workflow starts from elemental and prototype inputs, generates Y-Mn-B candidates through substitution and volume scaling, screens the generated structures with CGCNN formation-energy predictions, refines the low-energy set by MLIP relaxation and hull-energy ranking, and validates selected candidates with high-throughput DFT calculations. The output is an updated ternary convex hull and a set of predicted stable or metastable structures for subsequent structural, magnetic, phonon, and electronic analyses.

## 2.2 Candidate structure generation

A large pool of hypothetical structures was generated by substituting the target elements onto lattice sites of known crystal-structure prototypes. For the Y-Mn-B search, Y, Mn, and B were substituted into ternary crystal-structure prototypes. The prototype structure library contained 28,472 ternary structures extracted from the Materials Project [22], 1,013 unique magnetic structures from the Novomag database [23], and 7,068 additional unique structures from GNoME [24] after removing structural motifs similar to those already present in the Materials Project [22]. This procedure produced 36,553 structural templates.

For each structural template, 30 structures were generated by permuting the three elements over the crystallographic sites and applying five uniform volume scaling factors: 0.92, 0.96, 1.00, 1.04, and 1.08. The volume scaling accounts for size mismatch between the original prototype chemistry and the target Y-Mn-B chemistry. The final candidate pool contained 1,096,590 hypothetical ternary compounds. No DFT relaxation was performed before ML screening.

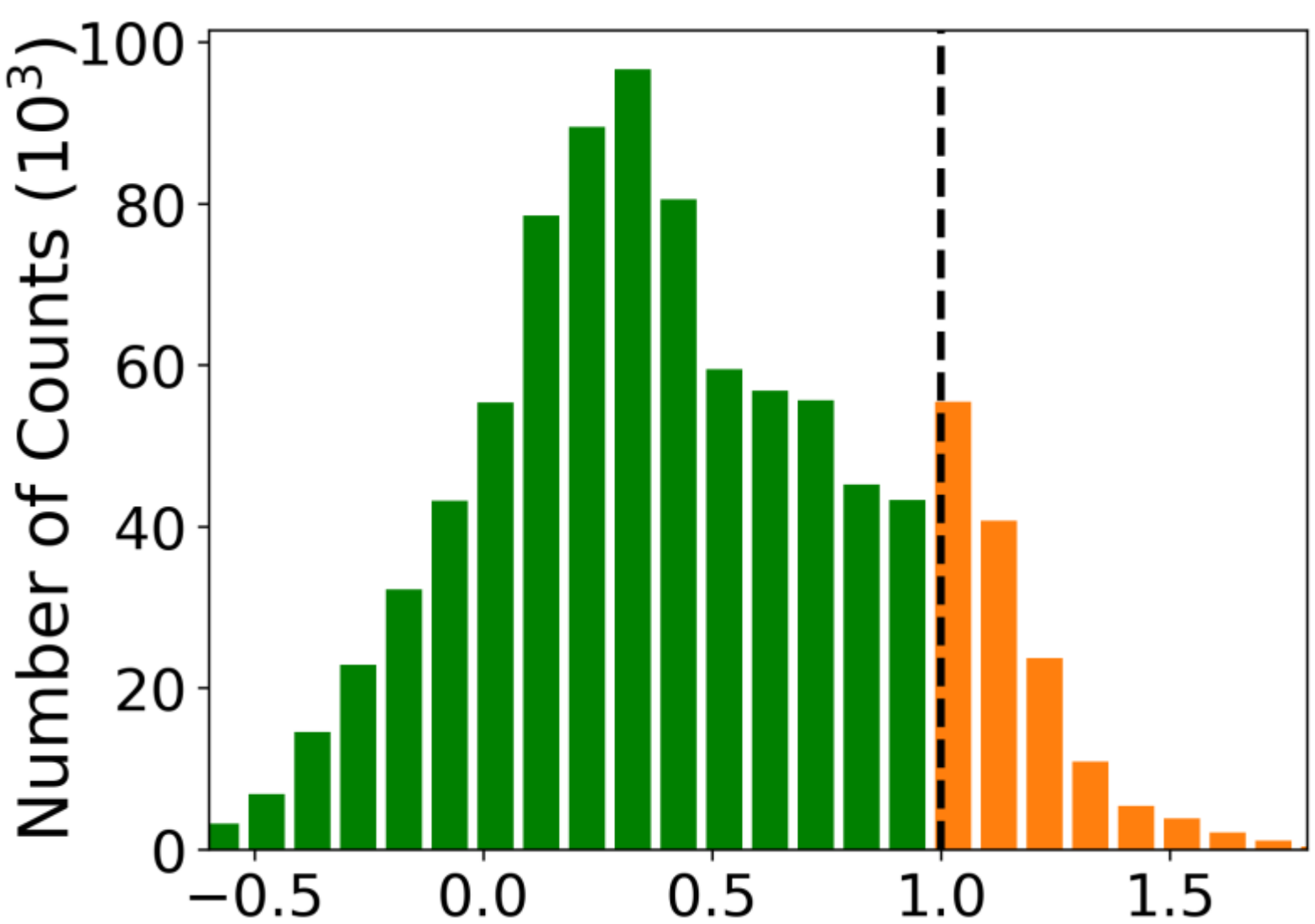

**Figure 2.** CGCNN-predicted formation-energy distribution of generated Y-Mn-B candidates. The histogram summarizes the first-stage CGCNN formation-energy predictions for the prototype-derived Y-Mn-B candidate pool. The dashed vertical line marks the 1.0 eV/atom screening threshold used to retain low predicted formation-energy structures for duplicate removal and structural filtering. After this redundancy removal, 10,050 structures were passed to MLIP relaxation and hull-energy sorting instead of sending the full generated pool directly to DFT.

## 2.3 CGCNN formation-energy screening

We first screened the candidate structures based on formation energies predicted by the crystal graph convolutional neural network (CGCNN) model [25]. CGCNN represents a crystal structure as a graph, with atoms as nodes and interatomic bonds as edges, enabling efficient learning of composition-structure-property relationships in inorganic crystals. We selected CGCNN for this first-stage screen because it provides a fast, well-established formation-energy surrogate for large pools of inorganic structures. At this stage, the goal is not to establish the final thermodynamic ranking, but to rapidly eliminate clearly unfavorable candidates before the more expensive relaxation and DFT validation steps.

The CGCNN model used in our exa-AMD package was trained on a broad and diverse dataset of 36,553 initial template structures, with the mean absolute error of 0.03 eV/atom for the formation energy prediction. Since this CGCNN model is trained for various crystalline structures and covers a wide range of chemical elements in the periodic table, it can be regarded as a "universal model" and thus can be adapted in our exa-AMD package to study a wide range of inorganic materials systems. The formation energy prediction of 1,096,590 structures takes 15 minutes to complete using a single GPU node with four NVIDIA A100 GPUs.

By applying the CGCNN model to the 1,096,590 hypothetical ternary compounds, structures with low predicted formation energies, $E_f$, were selected as promising candidates. In this work, we used a criterion of $E_f < 1.0$ eV/atom for this first down-selection. The CGCNN formation-energy distribution is shown in Fig. 2. A subsequent filtering step removed duplicate or structurally similar candidates, yielding 10,050 structures, approximately 0.92% of the original generated pool, for MLIP relaxation and hull-prioritized sorting described below.

## 2.4 MLIP relaxation and hull-prioritized sorting

The 10,050 down-selected structures were relaxed using FAIRChem's Universal Models for Atoms (UMA) [26], a machine-learning interatomic potential, together with the FIRE optimizer [27]. The resulting relaxed geometries were then used as the starting structures for subsequent DFT validation. The MLIP relaxations were executed in parallel within the exa-AMD workflow. Relaxation of the 10,050 selected structures can be completed in less than 120 minutes on a single node with four A100 GPUs.

After structural relaxation, the MLIP was used to predict the total energies, from which the formation energies and energies above the convex hull were estimated. The formation energy per atom for a ternary compound $Y_{\alpha}Mn_{\beta}B_{\gamma}$ was computed relative to the elemental reference phases:

$$E_f = [E(Y_{\alpha}Mn_{\beta}B_{\gamma}) - \alpha E(Y) - \beta E(Mn) - \gamma E(B)] / (\alpha + \beta + \gamma).$$

The total energy of each candidate structure was predicted with MLIP and converted to $E_f$ using the elemental reference energies defined above. The MLIP-estimated energy above hull $E_{\text{hull}}$, was then calculated relative to a reference convex hull constructed from DFT energies of the elemental, binary, and known stable ternary phases. In the present Y-Mn-B search, the MLIP-assisted hull sort selected 2,620 structures with $E_{\text{hull}} < 0.3$ eV/atom. This subset corresponds to 26% of the 10,050 structures that survived the CGCNN and redundancy filters and only 0.24% of the 1,096,590 generated structures. Thus, the MLIP step combines structural relaxation with a hull-aware triage, converting the list of structures screened by the formation energy into a prioritized set for DFT evaluation.

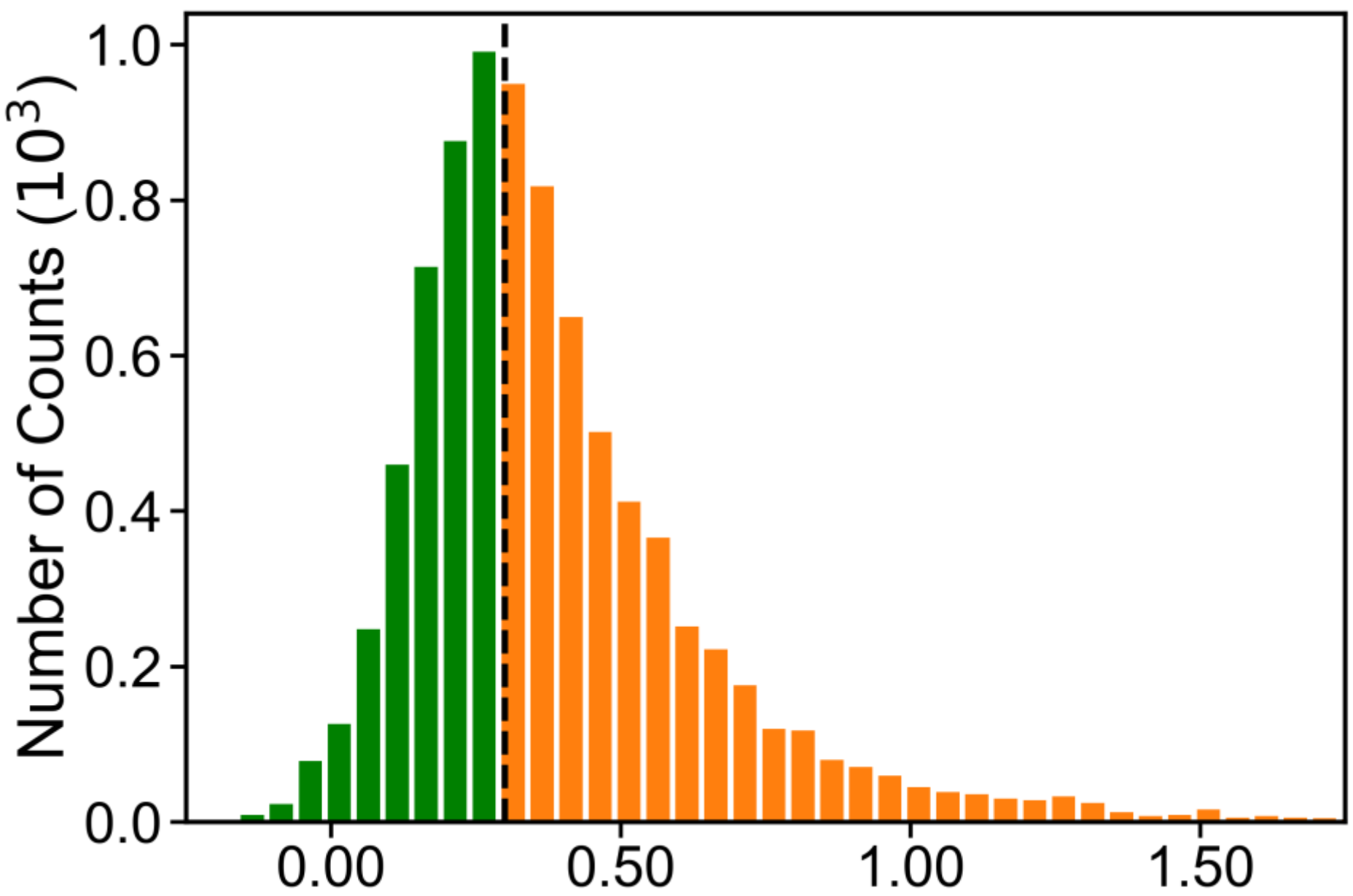

**Figure 3.** MLIP-assisted hull-energy sorting after CGCNN screening. The histogram shows the MLIP-estimated $E_{hull}$ distribution after relaxation of the 10,050 CGCNN-selected and redundancy-filtered structures. The dashed vertical line marks the 0.3 eV/atom cutoff used to select 2,620 structures for DFT-prioritized calculations, corresponding to 26% of the MLIP-relaxed set and 0.24% of the entire generated candidate pool.

## 2.5 First-principles calculations

First-principles calculations were performed using the Vienna Ab initio Simulation Package (VASP) [28-29]. The projector-augmented wave (PAW) method and the Perdew-Burke-Ernzerhof generalized-gradient approximation were used for the exchange-correlation functional [30-32]. The plane-wave energy cutoff was 520 eV. Brillouin-zone sampling used Monkhorst-Pack k-point grids with a density of $2\pi \times 0.025$ Å$^{-1}$ for structural relaxations [33]. Lattice vectors and atomic positions were fully relaxed until the residual force on each atom was less than 0.01 eV/Å.

The exa-AMD workflow parallelized the DFT relaxation stage across multiple CPU or GPU nodes. Depending on structure size and chemistry, DFT optimization of approximately 5,000 structures can be completed within one day using eight GPU nodes with 32 A100 GPUs. Individual run times depend on structure size, convergence behavior, and queue allocation.

After DFT relaxation, the convex hull was reconstructed from the DFT formation energies for the Y-Mn-B ternary system. A compound was considered stable at 0 K if it lays on the convex hull, $E_{hull} = 0$. Structures with $E_{hull} \leq 0.1$ eV/atom were treated as metastable candidates that may be experimentally accessible under non-equilibrium or finite-temperature synthesis conditions, following prior statistical analyses of experimentally reported compounds [34].

Magnetic moments were obtained from spin-polarized DFT calculations. The total cell magnetic moment, $M_{cell}$, was calculated by summing atomic magnetic moments, and the saturation magnetization was estimated as:

$$M_s = M_{cell} / V_{cell}.$$

The magnetic polarization was then computed as:

$J_s = \mu_0 M_s$.

For high-throughput screening, initial calculations assumed a collinear ferromagnetic state. For selected promising structures, additional antiferromagnetic, spiral, ferrimagnetic, or non-collinear configurations were tested to determine whether the ferromagnetic (FM) state remained energetically favorable. Noncollinear spin spiral calculations used the generalized-Bloch description of spiral magnetic order with PAW potential [35-36]. For $YMn_4B_4$, the spin spirals are sampled with rotations from 0 to 180° in 15-degree increment, where 0° and 180° correspond to FM and antiferromagnetic (AFM) states as two endpoints, respectively. This places the collinear FM and AFM states on a continuous spin spiral angles with non collinear magnetic states in between.

Curie temperatures were estimated from DFT magnetic-energy differences using a mean-field approximation to a Heisenberg-like exchange model. For a selected AFM configuration, the DFT energy difference was defined as $\Delta E_{cell} = E_{AFM} - E_{FM}$. Mapping this value to an effective exchange field $J_0$ requires knowing how many magnetic exchange pairs are reversed by the chosen AFM pattern. Each pair changed from parallel to antiparallel contributes an energy change of approximately $2J_{ij}S^2$, so a general mapping must be tied to the actual magnetic pattern. For the order-of-magnitude estimates reported here, we used the common fully reversed-neighbor shortcut, in which the effective exchange per magnetic atom is approximated as $J_0 = \Delta E_{cell}/N_m$, where $N_m$ is the number of magnetic atoms in the cell. The mean-field Curie temperature ($T_C$) was then estimated as:

$T_C = 2\, J_0 / (3\, k_B)$.

These $T_C$ values are estimated from mean-field exchange energies, which usually overestimate the real Curie temperature due to the neglection of higher order non-Heisenberg exchange interactions and spin fluctuations.

Magnetocrystalline anisotropy constants were evaluated for promising non-cubic structures using non-collinear DFT calculations with spin-orbit coupling. Self-consistent spin-polarized collinear calculations were first performed, followed by non-self-consistent SOC calculations with symmetry disabled. A denser k-point mesh of $2\pi \times 0.016$ Å$^{-1}$ and an electronic convergence criterion of $10^{-8}$ eV were used. Energies were calculated for magnetization oriented along the Cartesian (100), (010), and (001) directions, with additional directions such as (110), (011), and (101) considered for orthorhombic structures when needed. Positive $K_1$ denotes uniaxial anisotropy, whereas negative $K_1$ denotes easy-plane anisotropy under the sign convention used here. The magnetocrystalline anisotropy energy (MAE) was calculated using standard first-principles approach, in which the MAE was obtained from spin-orbit coupling (SOC)-induced energy differences between different magnetization directions [37-39].

Harmonic phonon calculations were used as a dynamical-stability check for selected low-energy structures. The phonon spectra were obtained from finite-displacement force calculations using Phonopy [40]. These calculations are reported separately from convex-hull stability because phonon stability tests the local curvature of a relaxed structure, whereas $E_{hull}$ tests the thermodynamic competition against decomposition products.

# 3. Results and discussions

## 3.1 Efficiency of MLIP-assisted discovery and updated Y-Mn-B hull

Navigating the Y-Mn-B search space presents a formidable combinatorial challenge. Conventional prototype-based substitution generates a vast number of chemically implausible structures, and genuinely viable candidates require substantial structural relaxation before true thermodynamic stability can be accurately assessed. Relying solely on ML-predicted formation energies to select unrelaxed structures for DFT calculations can miss promising candidates and lead to substantial computational inefficiency. The exa-AMD workflow circumvents this problem by employing a two-step approach: an initial CGCNN screening first eliminates a large portion of high-energy candidates from the 1.09 million generated Y-Mn-B structures, followed by a MLIP relaxation and $E_{hull}$-based prioritization step that further improves the computational efficiency.

Among the 1,096,590 hypothetical structures generated from database-derived ternary prototypes and volume-scaled site permutations, 10050 structures were passed to the FAIRChem UMA potential for relaxation. This MLIP relaxation step allows the structures to adapt to local Y–Mn–B bonding environments before DFT calculations. Figure 4a compares MLIP-estimated and DFT-calculated $E_{hull}$ for 1,537 structures for which both values were available. The mean absolute error over the complete set is 0.038 eV/atom. For the low energy regime of 192 structures with DFT-calculated $E_{hull} \leq 0.1$ eV/atom, the mean absolute error is 0.028 eV/atom.

To assess the efficiency and accuracy gain by introducing this MLIP relaxation step to estimate the relaxed $E_{hull}$, we examine the recovery percentage of stable and low-energy candidates with this MLIP step, compared to a pool established by CGCNN-only workflow. A total of 2,620 structures with $E_{hull} <$ 0.3 eV/atom were selected for DFT calculations to maintain coverage. Figure 4b shows that with the use of MLIP for relaxation, more than 95% of all DFT-validated stable and metastable candidates with $E_{hull} <$ 0.1 eV/atom are recovered within the first 500 selections, and more than 98% within the first 1,000. In contrast, CGCNN-only prioritization recovers only about 40% of those candidates after more than 2,600 selections. The added MLIP relaxation and hull-sorting layer significantly improve the accuracy and efficiency, compared to the original framework without these steps.

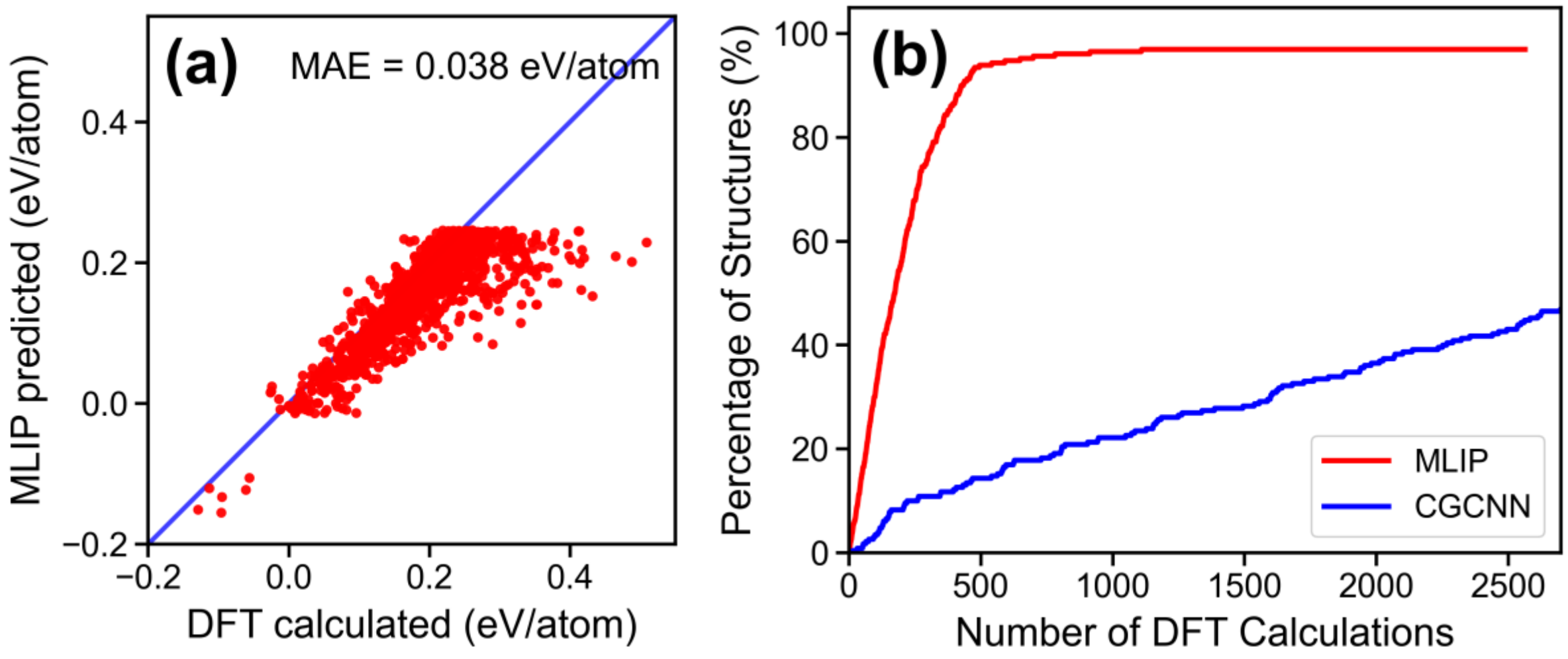

**Figure 4. Discovery advantage of the added MLIP relaxation and hull-sorting step. (a) Comparison between MLIP-predicted and DFT-calculated hull energies for representative low-energy Y-Mn-B candidates. The mean absolute error (MAE) is about 0.038 eV/atom. (b) Cumulative recovery of DFT-validated stable and metastable candidates with $E_{hull}$ < 0.1 eV/atom when the DFT calculations are prioritized by MLIP-estimated hull energy (red) compared with the original CGCNN-prioritized ordering (blue).**

The 2,620 structures selected by MLIP screening yielded 1,352 distinct structures after DFT relaxation and duplicate removal. Structures that did not reach electronic self-consistency or structural convergence were discarded from the final hull construction. We then recalculated formation energies and $E_{hull}$ values for the 1,352 DFT-relaxed ternary structures relative to the updated Y-Mn-B convex hull.

Prior to this work, the thermodynamic landscape of this system was largely uncharted, with only two thermodynamically stable ternary compounds, $Y_3MnB_7$ and $YMnB_4$. These two compounds are included in the existing convex hull in the Materials Project database, as illustrated in Figure 5(a). $Y_3MnB_7$ belongs to the $Er_3CrB_7$-type $R_3MB_7$ family of boron-rich rare-earth transition-metal borides. Experimental studies showed that synthesized $Y_3MnB_7$ is an orthorhombic *Cmcm* phase, and identified the $R_3MnB_7$ members with R = Y and Gd as isotypic with the $Er_3CrB_7$ structure [41-42]. In contrast to $Gd_3MnB_7$, for which ferrimagnetic ordering has been reported, band-structure calculations revealed that $Y_3MnB_7$ is not magnetically ordered [42]. $YMnB_4$ on the existing convex hull is also a known compound. It adopts the $YCrB_4$-type tetraboride in the *Pbam* space group, with Mn dimers or dimer-derived transition-metal units embedded in a B-rich framework [42-45]. Earlier work reported the synthesis and basic crystallographic properties of $YMnB_4$. Recent studies have renewed interest in the broader $RTB_4$ family because transition-metal dimers can produce quantum-dimer, conventional antiferromagnetic, or nonmagnetic behavior depending on the composition [43]. For $YMnB_4$, first-principles calculations have described a hard, predominantly covalent boride with semiconducting character, mechanical stability, and a computed band gap of approximately 0.25 eV [42,46]. More recent magnetic modeling places $YMnB_4$ among the conventional antiferromagnetic $RTB_4$ systems, with ferromagnetic Mn intradimer coupling but weaker three-dimensional interdimer interactions that favor antiferromagnetic ordering, giving a mean-field ordering temperature near 120 K [43].

Our ML-accelerated search results in a significant update of the ternary Y-Mn-B phase diagram, as shown in Fig. 5 (b). We identify four previously unreported stable Y-Mn-B compounds on the updated DFT hull, as shown by the red points in Fig. 5(b): $Y_2Mn_7B_7$, $Y_5(MnB_3)_2$, $Y(MnB_3)_2$, and $Y_2Mn_{25}B$. The related $YMn_4B_4$ phase lies close to the hull, with $E_{hull}$ = 10 meV/atom as shown in Table 1, and is the compact $R_{1+\varepsilon}T_4B_4$ chain approximant used for detailed magnetic investigation. Together, these stable and near-stable phases significantly update Y-Mn-B convex hull. In addition to the stable compounds, our workflow identified 61 unique metastable phases within 0.1 eV/atom of the updated hull (as shown in Supplemental Materials Table S1), which may also be experimentally accessible under appropriate synthesis conditions.

Among these phases, stable $Y_2Mn_7B_7$ and near-stable $YMn_4B_4$ are the most interesting since they are Mn analogs of the well-studied $R_{1+\varepsilon}Fe_4B_4$ chain family. In the Fe-based compounds, experiments describe the rare-earth and Fe-B substructures as incommensurate or commensurately approximated one-dimensional chain motifs, with ε typically near 0.10-0.15 depending on the rare-earth element and structural model [5,7-10].

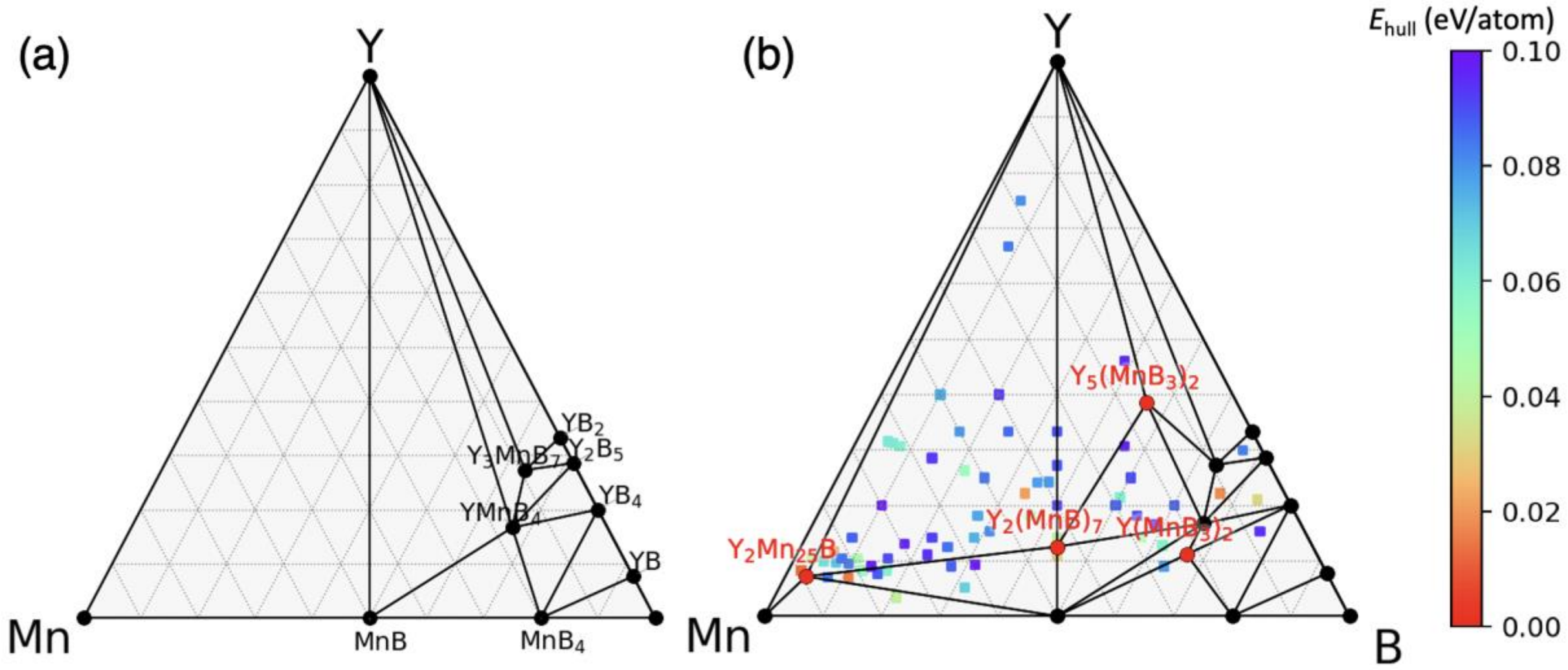

**Figure 5. Convex-hull update produced by the exa-AMD/DFT validation workflow. (a) Existing Y-Mn-B convex hull from the Materials Project [22]. Stable phases are shown as black points, including the experimentally synthesized reference phases $Y_3MnB_7$ (*Cmcm*) and $YMnB_4$ (*Pbam*). (b) Updated DFT convex hull from the present exa-AMD search, showing newly predicted stable $Y_2(MnB)_7$, $Y_5(MnB_3)_2$, $Y(MnB_3)_2$, and $Y_2Mn_{25}B$ phases as red points. The $YMn_4B_4$ phase is near stable, with $E_{hull}$ = 10 meV/atom above the updated hull.**

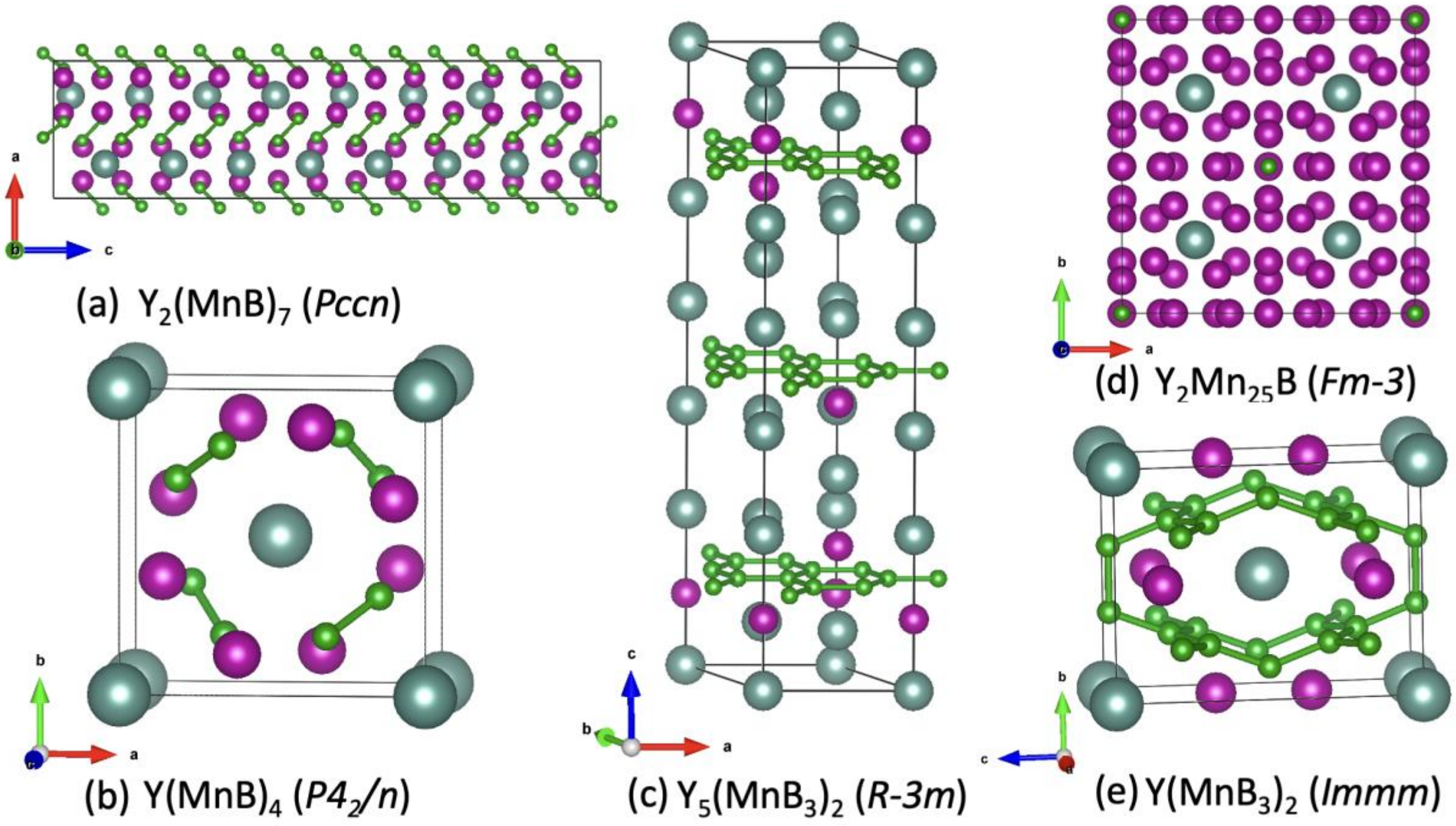

**Figure 6. Representative predicted stable and near-stable Y-Mn-B crystal structures. (a) $Y_2(MnB)_7$ (*Pccn*), (b) $Y(MnB)_4$ (*P4₂/n*), (c) $Y_5(MnB_3)_2$ (*R-3m*), (d) $Y_2Mn_{25}B$ (*Fm-3*), and (e) $Y(MnB_3)_2$ (*Immm*).**

$Y_2Mn_7B_7$ crystallizes in the orthorhombic *Pccn* space group and is best regarded as the 2-7-7 commensurate member of the $R_{1+\varepsilon}T_4B_4$ family, with ε = 1/7 = 0.143. The structure, as shown in Fig. 6(a), has lattice parameters of a = 7.0287 Å, b = 7.0288 Å, c = 27.6678 Å, with Z = 8 and an extended repeat along the c direction. For comparison, a single-crystal work on $Nd_{1.143}Fe_4B_4$ reported a related *Pccn* approximant with a = b = 7.130(3) Å and c = 27.358(9) Å [7], while the refinement of $Nd_{1+\varepsilon}Fe_4B_4$ described a = 7.117 Å with distinct Fe-B and rare-earth repeat lengths along c [8]. The Y-Mn-B chain phase therefore preserves the characteristic approximately 7 Å basal metric and approximately 3.9 Å transition-metal-boride repeat of the Fe-chain family, but with a slightly contracted basal plane and an elongated 2-7-7 repeat after replacing Nd/Fe by Y/Mn. Structurally, the Y atoms form a spacer substructure, whereas the Mn-B network provides the transition-metal-boride framework. The shortest Mn-Mn and Mn-B bonds are approximately 2.44 Å and 2.04 Å, respectively, placing Mn in a compact edge-sharing chain environment rather than in the dimerized $YMnB_4$ tetraboride environment as previously discussed.

$YMn_4B_4$ (Fig. 6(b)), is the 1-4-4 approximant of the family used as the prototype motif in a primitive cell. The optimized structure has a tetragonal *P4₂/n* cell with a = b = 7.0181 Å, c = 3.9240 Å. These parameters are close to the Fe-B subcell metric in $Nd_{1+\varepsilon}Fe_4B_4$, where $c_{Fe}$ = 3.897 Å was reported for the Nd member [8]. $YMn_4B_4$ is therefore the most compact crystallographic representation of the chain environment and is used below for electronic-structure and magnetic-property studies.

$Y_5(MnB_3)_2$ (Fig. 6(c)) in the *R-3m* space group is a boron-rich and comparatively Mn-dilute compound. It has lattice parameters a = 5.4432 Å, c = 23.0074 Å, with 3 formula units. Y is distributed over three sites with multiplicities 6, 6, and 3 along the rhombohedral axis. Mn occupies a single sixfold site; and B occupies an 18-fold site. The shortest Mn-Mn distance is about 3.94 Å, nearest Mn-B bonds are about 2.17 Å, and the shortest B-B distance is about 1.81 Å. The long Mn-Mn bonds and boron-rich coordination distinguish this phase from the chain compounds.

$Y_2Mn_{25}B$ (Fig. 6(d)) in the *Fm-3* space group represents the opposite structural limit. The conventional cubic cell has 4 formula units and a lattice parameter of 11.0260 Å, with Y on one 8-fold site, B on one 4-fold site, and Mn distributed over 48-, 48-, and 4-fold sites. The shortest Mn-Mn bond is approximately 2.22 Å, and the shortest Mn-B bond is approximately 2.30 Å. The structure is therefore dominated by 3D Mn clusters with dilute B and Y sites. This dense transition-metal framework is structurally reminiscent of the transition-metal-rich $RT_{13}$ and $NaZn_{13}$-type compounds, such as the $La(Fe,Al)_{13}$-based magnetocaloric materials, where magnetic behavior is highly sensitive to transition-metal connectivity and chemical substitution [47]. In $Y_2Mn_{25}B$, however, the B replacement of Mn introduces a much-complicated sublattice.

$Y(MnB_3)_2$ (Fig. 6(e)) is another boron-rich compound in the lower-symmetry *Immm* space group and exhibiting shorter Mn-Mn bonds. The unit cell has lattice parameters of a = 3.0913 Å, b = 6.3158 Å, c = 8.2514 Å, with two formula units. Y occupies one twofold high-symmetry site, Mn occupies one fourfold site, and B is split over two inequivalent sites with multiplicities 8 and 4. The shortest Mn-Mn bond is about 2.57 Å, nearest Mn-B bonds are about 2.06 Å, and the shortest B-B bond is about 1.73 Å.

We summarize the crystallographic information, $E_{hull}$ and calculated magnetic properties in Table 1.

**Table 1. Crystallographic, thermodynamic, and magnetic properties for the five new stable or near-stable Y-Mn-B compounds predicted in this work.**

| Compound | Space group | Z | Lattice parameters (Å) | $E_{hull}$ (meV/atom) | *Js* (T) | Magnetic ground state |
|---|---|---|---|---|---|---|
| $Y_2Mn_7B_7$ | *Pccn* | 8 | a = 7.0287, b = 7.0288, c = 27.6678 | 0 | 0.6 | FM |
| $YMn_4B_4$ | *P4₂/n* | 2 | a = b = 7.0181, c = 3.9240 | 10 | 0.6 | FM |
| $Y_5(MnB_3)_2$ | *R-3m* | 3 | a = b = 5.4432, c = 23.0074, γ = 120° | 0 | 0 | AFM |
| $Y(MnB_3)_2$ | *Immm* | 2 | a = 3.0913, b = 6.3158, c = 8.2514 | 0 | 0 | AFM |
| $Y_2Mn_{25}B$ | *Fm-3* | 4 | a = b = c = 11.0260 | 0 | 0.2 | FiM |

Thermodynamic stability is not by itself a proof of synthesizability. A critical additional criterion for evaluating a newly predicted crystal structure is its dynamical stability, which can be assessed using harmonic phonon calculations. A structure is considered dynamically stable at 0 K when no imaginary phonon branches appear across the sampled Brillouin-zone path. The calculated phonon dispersions for the predicted low-energy Y-Mn-B phases are shown in Supplementary Fig. S1 and indicate that these phases are dynamically stable within the harmonic finite-displacement approximations. We also carried out self-consistent spin-polarized DFT band-structure and density-of-states calculations for the other stable phases; these electronic structures are shown in Fig. S2.

As listed in Table 1, the boron-rich $Y(MnB_3)_2$ and $Y_5(MnB_3)_2$ entries favor AFM order. The dense-Mn-network in the $Y_2Mn_{25}B$ phase is identified as Ferrimagnetic, where Mn moments aligned in opposite directions but with different magnitudes, leading to a non-zero total moments. These compounds indicate that for complex magnetic ordering, low hull energy or high Mn content alone do not guarantee a robust

FM magnetic ground state. We will focus on two $R_{1+\varepsilon}T_4B_4$-type chain compounds, $YMn_4B_4$ and $Y_2Mn_7B_7$, where the crystal chemistry, electronic reconstruction, and magnetic ordering can be connected to the known Fe-based chain family.

## 3.2 Electronic and magnetic properties of the $Y_{1+\varepsilon}Mn_4B_4$ chain compounds

Previous work on the Fe-based $R_{1+\varepsilon}Fe_4B_4$ chain family established that this composite chain architecture is close to a magnetic instability. Mössbauer and magnetization studies found the Fe moments in the Fe-chain members to be strongly suppressed or essentially quenched, and long-range order in the Nd-based compounds appears only at low temperature and is governed primarily by the rare-earth sublattice rather than by an active Fe magnetic network [7,9–11]. Consistent with these observations, our DFT calculations for the Fe-chain compounds confirm that the Fe moments in these compounds are nearly extinguished, failing below 0.1 $\mu_B$ per Fe atom. In contrast, for structural analogs where Fe is replaced by Mn, $YMn_4B_4$ and $Y_2Mn_7B_7$, we find that they preserve massive local Mn moments of approximately 1.1 $\mu_B$ per atom within the ferromagnetic configuration. This stark difference illustrates that the 1D boride chain is not inherently magnetically inactive; rather, it is the transition-metal occupancy that dictates the magnetic properties.

To understand the microscopic origin of this difference, we compare spin-polarized band structures and densities of states (DOS) for the Mn- and Fe-based chain analogs in the FM configuration (Fig. 7). The comparison keeps the $R_{1+\varepsilon}T_4B_4$-type chain motif fixed within each structural pair, $Y(MnB)_4$ versus $Y(FeB)_4$ and $Y_2(MnB)_7$ versus $Y_2(FeB)_7$, so that the main variable is the transition-metal electron count. In the weakly magnetic Fe-chain references, the Fermi energy $E_F$ is far away from the strongest transition-metal 3*d* DOS peak and the majority/minority spin channels show only weak exchange splitting. Replacing Fe by Mn lowers the number of valence electrons and shifts $E_F$ toward a higher-DOS region of the Mn-derived 3*d* bands.

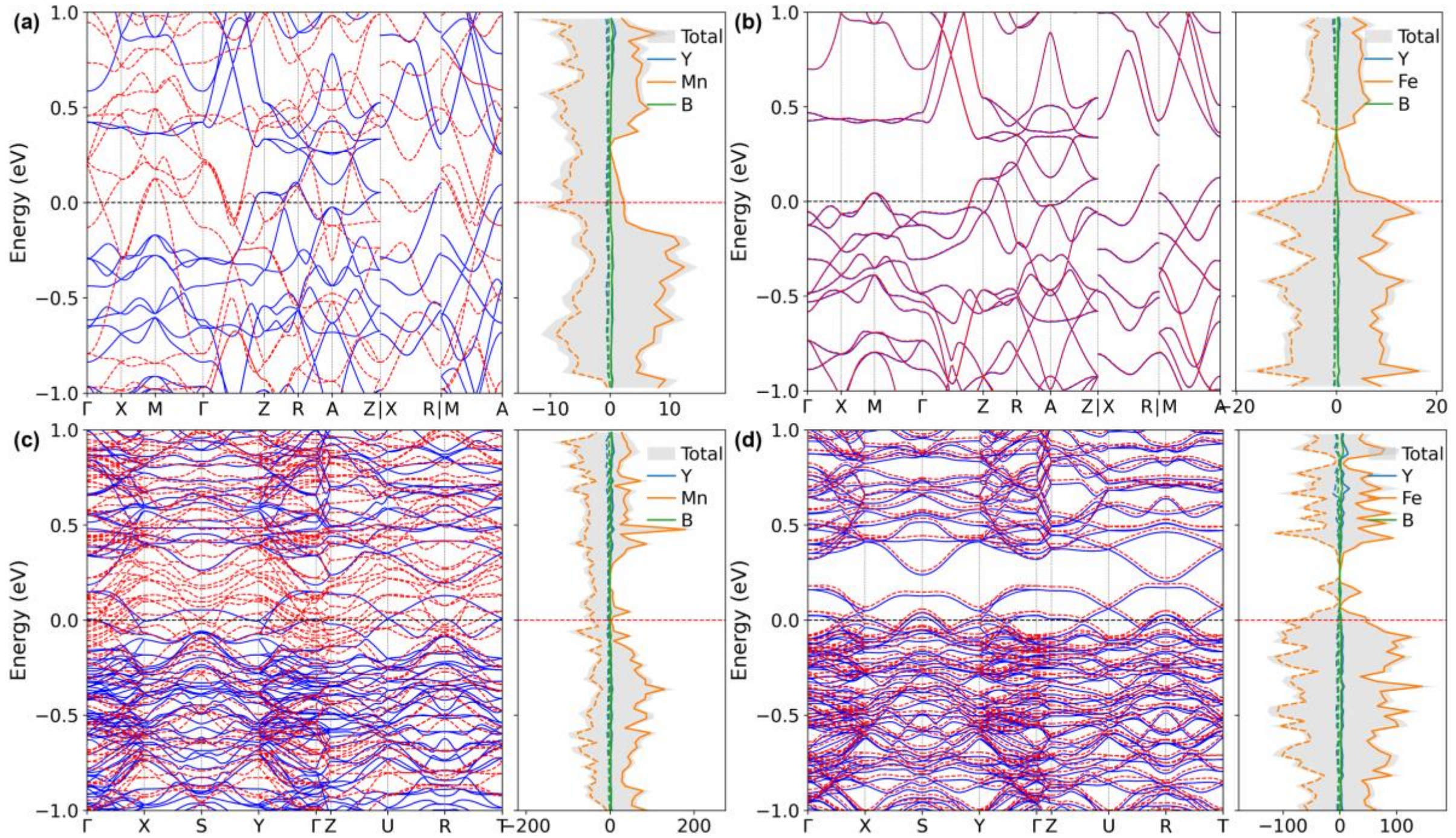

**Figure 7. Electronic-structure comparison of Fe- and Mn-based $Y_{1+\varepsilon}T_4B_4$ chain analogs. Spin-polarized band structures and densities of states for (a) $Y(MnB)_4$, (b) $Y(FeB)_4$, (c) $Y_2(MnB)_7$, and (d) $Y_2(FeB)_7$. The upper row compares the compact 1-4-4 Mn and Fe chain analogs, while the lower row compares the 2-7-7 Mn and Fe chain analogs.**

This electronic reconstruction reveals a Stoner-like instability: the accumulation of states at the Fermi level drives a pronounced exchange splitting, pushing the Mn-derived majority spin states downward relative to the minority spin states. Thus, the Mn substitution fundamentally alters the band occupancy and spin splitting, converting a non-magnetic Fe-boride chain structure into a Mn magnetic sublattice.

While local moments are substantially formed within the FM configuration, Mn-rich intermetallics are notoriously susceptible to competing FM, AFM, ferrimagnetic, and noncollinear states. For instance, Mn substitution in the archetypal $Nd_2Fe_{14}B$ permanent magnet matrix induces strong AFM coupling and significantly reduces the Curie temperature [12,13]. To determine the magnetic ground state of the Y-Mn-B chain compounds, we evaluated multiple collinear AFM configurations. Our calculations indicate a robust FM ground state. The lowest energy AFM configuration remain energetically unfavorable by approximately 29 meV/Mn for $Y_2Mn_7B_7$ and 10.8 meV/Mn for $YMn_4B_4$, compared to the FM configuration. Notably, the AFM states still support sizable local Mn moments (0.7 ~ 1.0 $\mu_B$), confirming that the underlying physical challenge in this framework is exchange selection between preformed moments rather than moment collapse.

To rule out the presence of intermediate noncollinear ground states, we performed a continuous noncollinear spin-spiral calculation for the compact $YMn_4B_4$ approximant (Fig. 8). The spin spiral was generated along the c direction and sampled rotation angles from the FM state at 0° to the AFM state at 180°. Along this sampled path, the total energy rises monotonically from the FM endpoint and reaches approximately 29 meV/atom at 180°, without a local minimum with an intermediate angle. For the spin spiral q vectors investigated here, there is no sign of a lower energy noncollinear state between the two collinear limits, which confirm the collinear FM state as the magnetic ground state.

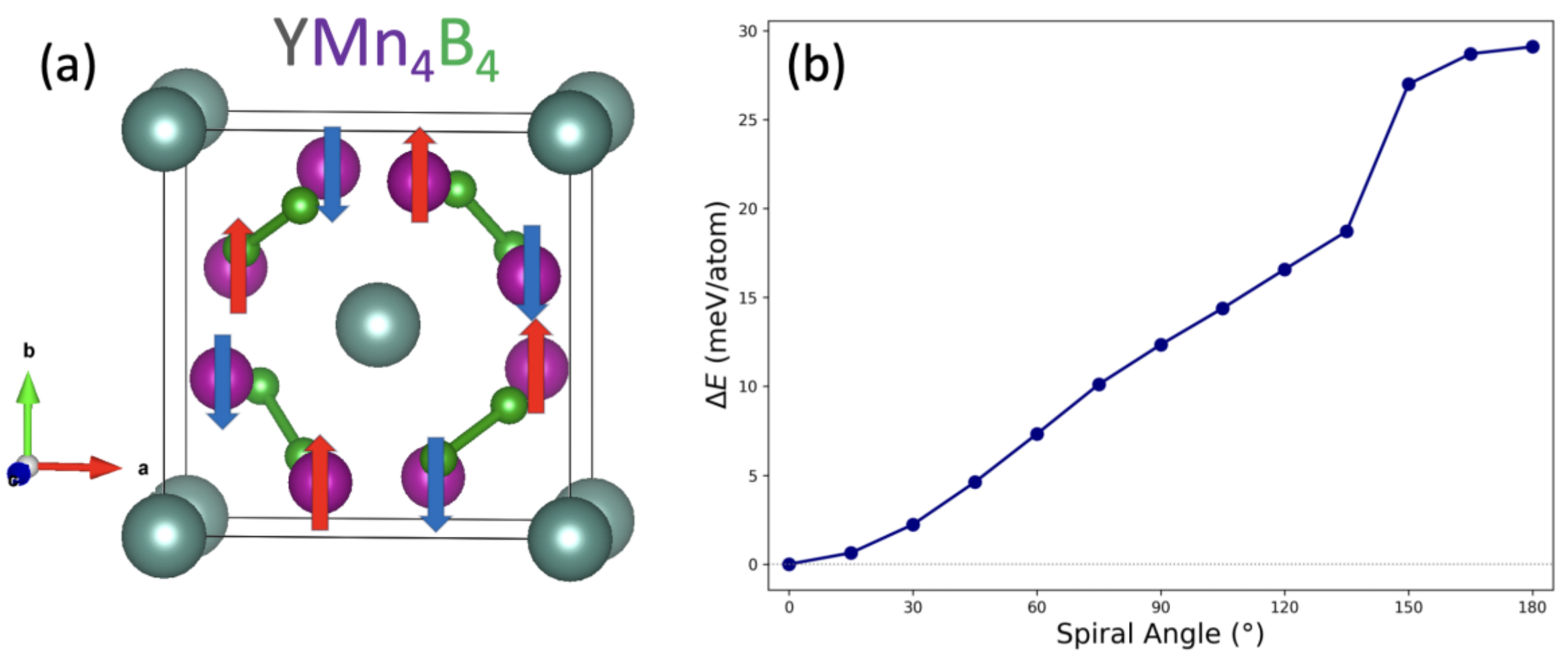

**Figure 8. Noncollinear spin-spiral calculations for $YMn_4B_4$. (a) Schematic spin arrangement used to test competing Mn-moment alignment in the compact $YMn_4B_4$ chain approximant. (b) Noncollinear spin-spiral energy curve from the FM endpoint at 0° to the AFM endpoint at 180°. Energies are reported relative to the FM state.**

Using the mean-field approximation (MFA) described in the Methods section, we derive comparative Curie temperature ($T_C$) estimates. With an energy difference of 10 meV/Mn, we predict $T_C^{MFA} \approx 77$ K for $YMn_4B_4$. For $Y_2Mn_7B_7$, the larger FM and AFM energy difference gives rise to a higher $T_C^{MFA} \approx 210$ K. We emphasize that these MFA values represent relative exchange-energy scales rather than precise critical temperatures, as long-range magnetic ordering in quasi-one-dimensional systems is strongly suppressed by spin fluctuations neglected in mean-field theory.

This comparison suggests that the longer 2-7-7 chain compound may stabilize magnetic ordering more effectively than the compact 1-4-4 approximant. Previous work on $R_{1+\varepsilon}Fe_4B_4$ compounds mainly establishes structural modulation, rare-earth/transition-metal substructure mismatch, and weak Fe-derived magnetism, rather than a general rule that composite chain structures raise $T_C$ [7,9-11]. We conclude that the 2-7-7 approximant slightly changes the Mn-Mn and Mn-B bonding orientations along z direction, compared to $YMn_4B_4$, and those changes correlate with a larger calculated AFM-FM energy difference.

Our spin-orbit coupling (SOC) calculations reveal an in-plane magnetocrystalline anisotropy for $YMn_4B_4$ and $Y_2Mn_7B_7$, with $K_1 \sim -1.73$ and $\sim -1.34$ MJ/m$^3$, respectively. The easy-plane anisotropy is consistent with the absence of localized rare-earth 4*f* crystal-field anisotropy: removing localized rare-earth 4*f* moments removes the anisotropy that contributes to much of the uniaxial hardness in rare-earth magnets [1-4]. The remaining anisotropy originates from the transition-metal-boride network and is evaluated through standard SOC total-energy differences between magnetization directions [37-39].

To evaluate the structural flexibility and magnetic tunability of this framework, we mapped the pseudo-binary substitution line $Y_2(Mn_xFe_{1-x})_7B_7$ (Fig. 9). The shallow thermodynamic penalty for ordered Fe/Mn mixing across the sampled substitution space, together with the higher formation energies of the sampled random configurations, indicates strong site preferences and a tendency toward chemical ordering, while also suggesting robust tolerance for alloying within the $R_{1+\varepsilon}T_4B_4$-type chain.

Remarkably, the calculated magnetic polarization $J_s$ increases nearly monotonically with Mn fraction, peaking at 0.553 T for pure $Y_2Mn_7B_7$. This trend shows that Mn incorporation into this architecture does not reproduce the magnetization collapse observed for Mn substitution in 2-14-1 Fe borides [12,13]. Instead, this line substitution provides a pathway for experimental efforts by manipulating the composition for tuning magnetic polarization while maintaining low calculated hull energies.

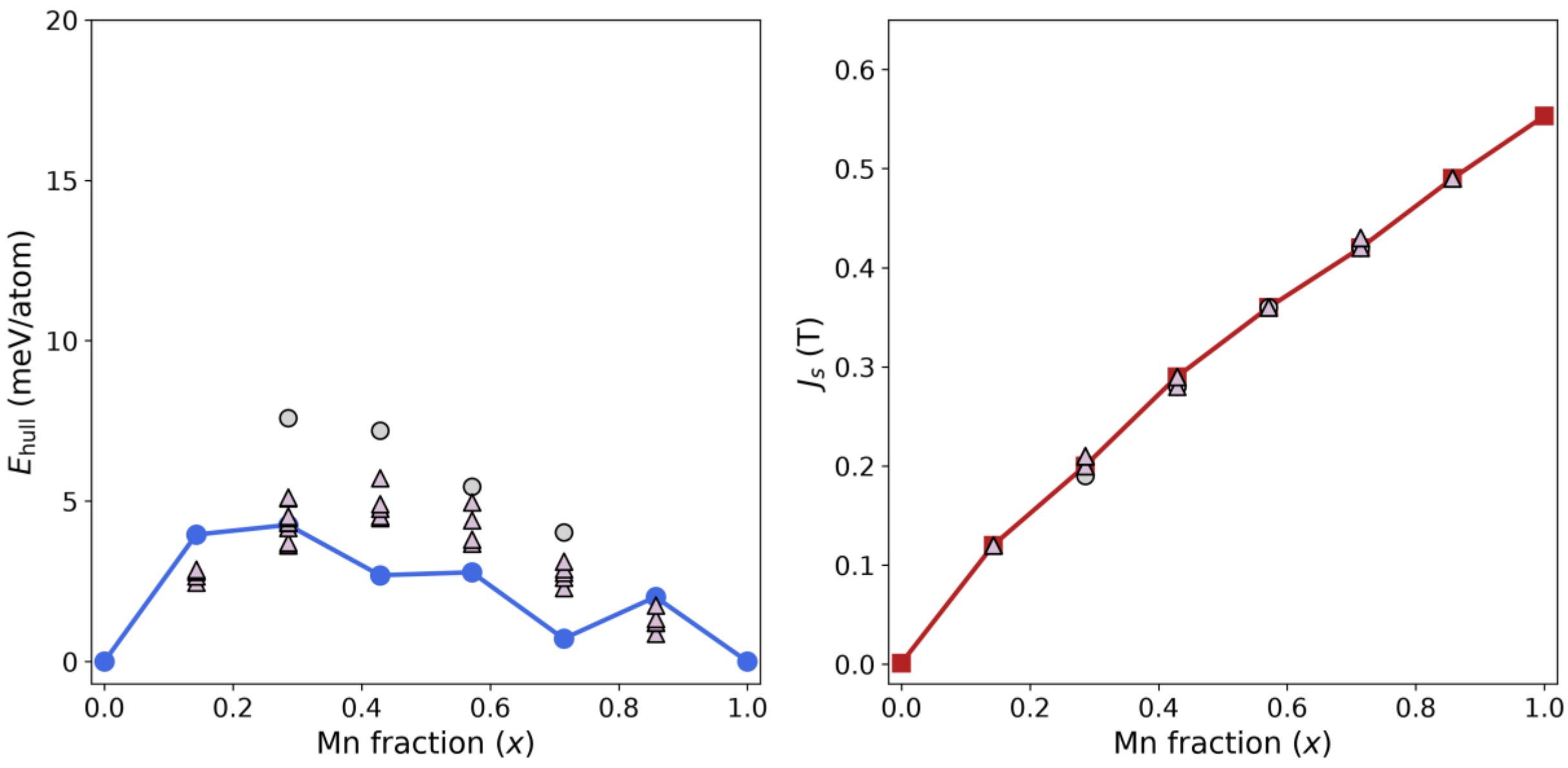

**Figure 9. Stability and magnetic polarization along the $Y_2(Mn_xFe_{1-x})_7B_7$ substitution line. $E_{hull}$ and magnetic polarization $J_s$ are plotted as a function of Mn fraction $x$ in $Y_2(Mn_xFe_{1-x})_7B_7$. Connected blue circles and red squares denote $E_{hull}$ and $J_s$, respectively, for the selected ordered composition series. Unconnected gray circles denote additional ordered configurations, whereas gray triangles denote random Fe/Mn arrangements.**

The MLIP-assisted exa-AMD workflow significantly accelerate the materials discovery, identifying new low energy Y-Mn-B phases with promising properties. The remaining exploration for high-performance permanent magnets, specifically strong uniaxial anisotropy and elevated transition temperatures, can be further investigated in future work by substituting Y with magnetic rare-earth elements such as Nd or Sm, which introduce critical 4*f* anisotropy and supplementary 3*d*-4*f* exchange pathways.

# 4. Summary

In this study, we applied the exa-AMD machine-learning workflow to systematically explore the non-4*f* Y-Mn-B ternary phase space. By integrating CGCNN formation-energy screening, MLIP relaxation, hull-prioritized sorting, and DFT calculations, the workflow reduced a 1,096,590-structure prototype pool to a tractable set of first-principles candidates while preserving the stable and low-energy structures needed to update the ternary hull. The MLIP relaxation and hull-sorting step is central to this efficiency because it reorders candidates after local structural relaxation rather than relying only on formation-energy predictions on unrelaxed structures.

The accelerated search identifies four newly predicted stable Y-Mn-B phases, $Y_2Mn_7B_7$, $Y_5(MnB_3)_2$, $Y(MnB_3)_2$, and $Y_2Mn_{25}B$, and near-stable $YMn_4B_4$, along with 61 metastable phases where $E_{hull}$ is within 100 meV/atom. The 5 thermodynamically favorable compounds are calculated to be also dynamically stable and show unique magnetic properties. $Y(MnB_3)_2$ and $Y_5(MnB_3)_2$ are boron-rich frameworks with different Mn-Mn bonds, while $Y_2Mn_{25}B$ is a dense-Mn-network compound. $Y(MnB_3)_2$ favors an AFM magnetic ground state, $Y_5(MnB_3)_2$ shows very small magnetization.

The predicted $Y_2Mn_7B_7/YMn_4B_4$ belongs to the $R_{1+\varepsilon}T_4B_4$-type structural family, which is previously known for crystallographic complexity and severely suppressed Fe magnetism. In the Mn-chain compounds, the magnetic properties are significantly different from the previously synthesized $R_{1+\varepsilon}Fe_4B_4$ incommensurate/composite chain family. The Mn versions retain sizable local moments of approximately 1.1 $\mu_B$/Mn and favor FM as the magnetic ground state. The reduced number of valence electron shifts $E_F$ into a high-DOS region dominated by Mn-derived 3*d* states, generating substantial exchange splitting and restoring a robust transition-metal moment. This is also in sharp contrast to the AFM widely observed when Mn is introduced into the 2-14-1 boride framework [12,13].

The calculated magnetic ordering temperatures remain modest at the mean-field level (below room temperature), and the SOC calculations indicate in-plane anisotropy. Although the predicted Y-Mn-B compounds are stable and exhibit substantial magnetization, replacing Y by anisotropic magnetic rare earths such as Nd or Sm may further improve their magnetic properties. Future studies will make use of advanced first-principles methods to examine whether the inclusion of rare-earth 4*f* anisotropy and additional 3*d*-4*f* exchange can preserve and enhance the phase stability and Mn-derived transition-metal magnetization, while increasing the Curie temperature and magnetic anisotropy.

## Data and code availability

The improved exa-AMD framework is available on GitHub: https://github.com/ml-AMD/exa-amd/. The data leading to the findings in this paper will be available on Zenodo upon publication.

## Acknowledgments

Work at Ames National Laboratory and Los Alamos National Laboratory was supported by the U.S. Department of Energy (DOE), Office of Science, Basic Energy Sciences, Materials Science and Engineering Division through the Computational Material Science Center program. Ames National Laboratory is operated for the U.S. DOE by Iowa State University under contract # DE-AC02-07CH11358. Los Alamos National Laboratory is operated by Triad National Security, LLC, for the National Nuclear Security Administration of U.S. Department of Energy under Contract No. 89233218CNA000001. This research used resources of the National Energy Research Scientific Computing Center (NERSC), a DOE Office of Science User Facility supported under Contract No. DE-AC02-05CH11231.